\documentclass[preprint,preprintnumbers,superscriptaddress,amsmath]{revtex4}
\usepackage{url}
\usepackage{epsfig}
\usepackage{graphicx,color}
\usepackage{epstopdf}
\usepackage[psamsfonts]{amssymb}
\usepackage{amsmath}
\usepackage{indentfirst}
\usepackage{ulem}

\usepackage{url}

\newcommand{\be}{\begin{equation}}
\newcommand{\ee}{\end{equation}}
\newcommand{\ba}{\begin{eqnarray}}
\newcommand{\ea}{\end{eqnarray}}

\begin{document}

\title{Quantification of deviations from rationality\\ with heavy-tails in human dynamics}

\author{T. Maillart}
\email{tmaillart@ethz.ch}
\affiliation{Department of Management, Technology and Economics,
ETH Zurich, Kreuzplatz 5, CH-8032 Zurich, Switzerland}

\author{D. Sornette}
\affiliation{Department of Management, Technology and Economics,
ETH Zurich, Kreuzplatz 5, CH-8032 Zurich, Switzerland}

\author{S. Frei}
\affiliation{Computer Engineering and Networks Lab (TIK), ETH Zurich, 8001 Zurich, Switzerland}

\author{T. Duebendorfer}
\affiliation{Google Switzerland GmbH, 8002 Zurich, Switzerland} 

\author{A. Saichev}
\affiliation{Department of Management, Technology and Economics,
ETH Zurich, Kreuzplatz 5, CH-8032 Zurich, Switzerland}

\begin{abstract}
The dynamics of technological, economic and social phenomena is
controlled by how humans organize their daily tasks in response to
both endogenous and exogenous stimulations. 
Queueing theory is believed to provide a
generic answer to account for the often observed power-law distributions of waiting
times before a task is fulfilled.
However, the general validity of the power law and
the nature of other regimes remain unsettled.
Using anonymized data collected by Google at the World Wide Web level, we 
identify the existence of several additional regimes characterizing 
the time required for a population of Internet users to execute a given task after receiving a message.
Depending on the under- or over-utilization of time by the population of users and the
strength of their response to perturbations,  the pure power law 
is found to be coextensive with an exponential regime (tasks are performed
without too much delay) and with a crossover to an asymptotic plateau (some tasks are never performed). 
The characterization of the availability and efficiency of humans on their actions revealed by our study
have important consequences to understand human decision-making, 
optimal designs of policies such as for Internet security,
with spillovers to collective behaviors, crowds dynamics,
and social epidemics.
\end{abstract}

\date{\today}

\maketitle

\section{Introduction}

Consider a typical individual, who is subjected to a flow of sollicitations. In the presence of time, energy, regulatory, social and monetary influences and constraints, such an individual must set priorities that will probably trigger a delay in the execution of task related to one message. 
Recent studies of various social systems have established the remarkable fact that
the distribution $Q(t)$ of waiting times between the presentation of the message and the ensuing action
has a power law asymptotic of the form 
$Q(t) \sim 1/t^\alpha$, with an exponent $\alpha$ often found smaller than $2$. Examples include the distribution of waiting time until a message in answered in email \cite{Eck} and mail \cite{Oliveira_Bara} correspondence, and in other human activity patterns, like web browsing, library visits, or stock trading \cite{Vasquez_et_al_06}.

A related measure concerns the response of an individual,
social or economic system, measured by a rate $R(t)$ of activity, following a perturbation,
an announcement or a shock.  
When the message is delivered directly and simultaneously to a large number of actors,
the measure of activity in this population can be shown to be proportional
to the distribution $Q(t)$ of waiting times (in the absence of word-of-mouth effects) \cite{craneSchSor2009},
or nonlinearly related to $Q(t)$ (in the presence of social influences) \cite{SaiSorthetagen10}.
Such measures of activity following a shock have also been documented 
to decay extremely slowly, approximately
like a power law of the time counted from the shock \cite{sornette2005origins}. For instance, measures
of media coverage after a large geo-political event 
(such as the assassination of the Prime
Minister of India on October 31, 1984, the coup against Gorbachev in
August 1990 or the terrorist attacks on September 11, 2001) decay approximately
as a power law of time since the occurrence of the event \cite{roehner2004news}. The rate of downloads of papers or book sales from a website after a media coverage also follows a power law decay \cite{johansen2000internaut,johansen2001internaut,sornette2004amazon,deschatres2005amazon}.
The relaxation of financial volatility after a peak follows
a power law, with different regimes \cite{VolMRW}. An alternative measure 
of the number of volatility peaks above some threshold also have been found
to exhibit this power law decay \cite{Lillo-Mantegna-Omori},
called the Omori law in analogy with a similar behavior of the rate of aftershocks following a large earthquake
The study of the rate of video views on YouTube has confirmed the existence of 
power law decay after peaks associated with media exposure  \cite{cranesorYouTube}, on a very large database of millions of videos. The dynamics of visitations of a major news portal  \cite{Dezso_et_al_06}, and
the decay of popularity of Internet blog posts  \cite{Leskovec_et_al_07} also follow the same 
power law decay. 

The following section 2 presents an economic model of priority queueing processes, whose
main point is to rationalize the empirical observation that systems are most often than not
found to operated close to the critical point of the priority queueing model. The attraction
to the critical point is explained in terms of the competition between improving the utility
of time allocation and the cost of time management. A slight extension of the priority
queueing model allows to account for behavioral biases and heterogeneity of agents. 
Section 3 then presents the data set and the results of the calibration of the empirical
data on browser updates by the model. Section 4 concludes. The appendix provides
relevant information on the data collection process.

\section{Economic model of priority queueing processes}

\subsection{Introduction to priority queueing models}

Such power laws have been rationalized by ``priority queueing'' models 
\cite{cobham1954,Barabasi_Nature05,grinstein2006,grinstein2008}, 
in which tasks enter the queue following some stochastic process  and are addressed in order of their priority. The key control parameter of these priority models is the ``time-deficit'' parameter  $\beta$, defined as the difference between the average time $\langle \tau \rangle$ needed to complete 
a task and the average time  interval $\langle \eta \rangle$ between task arrivals. 

For $\beta<0$, all tasks are  eventually completed while for $\beta>0$, there is a strictly positive probability for the low priority tasks never to be done. The value $\beta=0$ is a critical point of the  theory for which the cumulative distribution of waiting times exhibits an exact asymptotic power law tail $Q_{\beta=0}(t)\sim 1/t^\alpha$ with exponent $\alpha =0.5$. 
In the extremal version where a task is singled out with the lowest priority \cite{SaiSorProcras}, deviations of $\beta$
from $0$ can be classified into two opposite regimes:
for $\beta <0$, $Q(t)$ develops an exponential tail; for $\beta>0$, $Q(t)$ crosses over to a plateau at $0<Q_{\infty}<1$, equal to the probability that the low-priority task is never completed. 

Priority queueing models assume an exogenously determined mapping between the in-flow of tasks and their priorities, which  determine the ordering of their fulfillment.  In reality, there are certain tasks that may modify the whole task-solving process. 
We model this phenomenon by considering that the time-deficit parameter $\beta(t)$ is actually a dynamic variable, determined from the interplay between the incentive
for agents to maximize the utility of their time and the costs of adaptation to the stochastic flow of tasks. 


\subsection{Mapping of priority queueing onto consumption economic theory}

In general, individuals tend to maximize their utility given their available resources (which include the issue of time allocation). However, the essential difference between the standard consumption optimization and tasks allocation is that time is a limited and non storable resource. We proceed by analogy with the
standard utility theory of decision making, in which a representative agent
 has to optimize her consumption flow over her lifetime discounted by 
 her pure rate of time preference, given the intertemporal budget constraint
 that the present value of future consumption equals the present value of 
 future income \cite{Gollier}.  Let us now consider the following analogy:
 {\footnotesize
 \ba
{\rm available ~wealth} ~W   & \to &  {\rm available ~time} ~T  \\
{\rm consumption~ spends}~C~{\rm units ~of~wealth}  & \to &  {\rm solving~a~task~consumes~time} ~\tau
\nonumber \\
{\rm utility~from~consuming~} C:~ U(C) & \to &  {\rm gain~from~solving~a~task~taking~time}~ \tau:  ~U(\tau) \\
{\rm total ~utility~of~consuming~over~}T ~{\rm periods}: & \to &  {\rm total~utility~ from~ solving}~N ~{\rm tasks ~during~ time}~T~~~\\
U = \sum_{t=1}^T u(c_t) e^{-\rho t} dt  & \to &   U = \sum_{i=1}^N  u(\pi_i)  \\
{\rm budget~ constraint} ~\sum_{t=1}^T u(c_t) e^{-r t} dt  =\sum_{t=1}^T y_t e^{-r t} dt     & \to & 
 {\rm time~ budget} ~\sum_{t=1}^T  \tau_i  \leq T  ~.  
 \label{hbgw} 
 \ea
}
Here, $\rho$ is the pure rate of time preference and $r$ is the market interest rate.
In the standard program on the left, the rational agent aims at maximizing $U = \sum_{t=1}^T u(c_t) e^{-\rho t} dt $
subject to the budget constraint written on the left side of (\ref{hbgw}). In the program on the right, 
a rational agent attempts to maximize her utility $U = \sum_{i=1}^N  u(\pi_i)$ derived from performing the tasks
or occupations during a fixed time interval $T$,  subject to the condition that the total time spent 
on these tasks  ($\sum_{i=1}^N  \tau_i$)
cannot be larger than $T$.

\subsection{Optimal time consumption leads to convergence to the critical point $\beta \to 0$}

The essential difference between the optimization program on the right compared
to the standard consumption optimization on the left is that time is non-storable. Unused time is lost forever.
Thus, the rational agent will try to find new occupations so as to increase the number of tasks in each time interval available to her.
If she finds that her $N$ tasks performed during time $T$ consume $\sum_{i=1}^N  \tau_i$ which is less than $T$,
she will search for other tasks to increase her utility. This can take for instance the form
of education (the analog of investing for future income $y_t$) which provides novel opportunities of tasks 
in the future that will increase her future utility. She may also spend time exploring new opportunities, 
which is the analog of diversification of wealth investments for future consumption.
In this way, over the long run, these strategies will make the agent tend to increase the utilization $\kappa$ of her time defined by
\be
\kappa = {1 \over T} \sum_{i=1}^{N(T)}  \tau_i~.
\ee
The time budget constraint (\ref{hbgw}) then reads $\kappa \leq 1$.

With the definition of the time deficit parameter $\beta := \langle \tau \rangle - \langle \eta \rangle$,
in terms of the difference between the average of the duration $\langle \tau \rangle$ of a task and 
the average waiting time $\langle \eta \rangle$ between task arrivals, we have 
\be
\langle \kappa \rangle =  {1 \over T} \sum_{i=1}^{N(T)}  \langle \tau_i \rangle = {N \over T}  \langle \tau \rangle 
= {\langle \tau \rangle \over \langle \eta \rangle} = 1 + {\beta \over  \langle \eta \rangle}~,~~~{\rm for} ~\kappa <1~ (\beta <0)~,
\ee
where we have used that $ \langle \eta \rangle = T/N$ in the regime $\kappa \leq 1$. 
A negative time deficit parameter $\beta <0$ corresponds to a sub-optimal utilization ($\kappa <1$). 

As the agent increases $N$ for a fixed $T$, the utilization $\kappa$
will eventually reach the upper bound $1$ associated with the 
time budget constraint (\ref{hbgw}). As a consequence, the agent will not be able to increase her total utility anymore
even by increasing $N$ further, since only a subset of the $N$ available tasks will be performed in the allotted time $T$.
Worse, in this case, one needs to include a new term in the total utility of the agent, in order to 
account for the cost of task management that appears when the total burden of tasks increases
above the time available to solve them. Indeed, in this case, the agent who faces more tasks or occupations
that she can perform has to choose among all available tasks a subset that she will be
able to address in her available time $T$. This 
takes the form of a cost, either just in the form of time used to manage and prioritize her tasks
and of other resources spent to perform this management. Then, since the average utility
${\rm E}[U] = {\rm E}[ \sum_{i=1}^N  u(\tau_i)]$ saturates to a constant when $\sum_{t=1}^N  \tau_i$ reaches
$T$, the total utility which includes the management cost when $\sum_{t=1}^T  \tau_i  > T$ becomes
a decreasing function of the number of tasks. 

Therefore, in order to reduce waste of time (resp. overstress due to accumulation of pending tasks), a rational agent in a stationary set-up will tend to adjust her time-deficit $\beta$ as close to zero as is possible, therefore maximizing her utility. 
As for  geographies of cities or economies of firms \cite{MalSaiSor10,MalSaiSorbook}, 
the critical power law regime with $\alpha=0.5$ resulting from $\beta=0$, can be interpreted as the result of optimal allocation of resources that ensures the maximum sustainable use of time. 

We summarize the optimization process leading to the convergence to the critical point $\beta =0$ as follows.
\begin{itemize}
\item for $\kappa <1$, the total utility is less than optimal and the agent 
will increase $N/T$ by finding new tasks or occupations,  in order to increase her utility;
\item when $\kappa$ reaches $1$ and $\sum_{t=1}^N  \tau_i > T$, her utility becomes
a decreasing function of $N/T$ due to management costs.
\item Hence, the optimal number of tasks is such that $T/N :=  \langle \eta \rangle = \langle \tau \rangle$,
for which $\beta =0$ and the utilization $\kappa$ reaches marginally its upper bound $1$. This boundary
value for $\kappa$ corresponds to the critical value $\beta=0$ of the priority queueing theory, which is derived
here as the fixed attractive point of a utility maximizing agent under general conditions.
\end{itemize}

\subsection{Extension to account for behavioral biases and heterogeneity}

However, humans are no perfect time utility maximizers and suffer from cognitive biases as well as emotional quirks. 
This can result in departures from the optimum $\beta=0$, due to imperfect anticipations of new tasks
and changes in the solving rate.  Fig. 1 presents several possible scenarios on how the
arrival of a given task may trigger a perturbation, thus changing $\beta(t)$ and the response $Q(t)$. 
We capture these effects by modeling the dynamics of $\beta$ by a mean-reverting process around some attractive point $\overline{\beta}$ close to the critical value $\beta=0$.  

The simplest 
mean-reverting dynamics is the Ornstein-Uhlenbeck process \cite{Risken}, which predicts that the distribution of $\beta$'s estimated over a large population of similar agents is a Gaussian law centered on $\overline{\beta}$ with a standard deviation (std) $\beta_{0}$ determined by the efficiency with which agents optimize their use of time. The parameter $\overline{\beta}$ quantifies the intrinsic propensity of the population of agents
to under-use ($\overline{\beta}<0$) or over-stretch  ($\overline{\beta}>0$) their time. The std $\beta_{0}$
is proportional to the speed with which the agents adapt to changing conditions. Large $\beta_0$'s correspond
to a diffuse and weak response to perturbations and changing conditions of incoming tasks. 
Under these conditions, the population of agents can be parameterized by
 the Gaussian distribution $\psi (\beta; \overline{\beta}, \beta_{0})$  
 \begin{equation}
\psi (\beta; \overline{\beta}, \beta_{0}) = \frac{1}{ \sqrt{ 2\pi \beta_{0}}} \textrm{exp}\left(-\frac{(\beta-\overline{\beta})^{2}}{2\beta_{0}^{2}} \right)~.
 \label{gaussian}
\end{equation}

In the presence of such a distribution of time deficit parameters $\beta$, assuming independence between individuals with different $\beta$'s, the survival distribution $\overline{Q}(t)$ of the waiting times until completion of the target task is simply
the average of the one-person survival distribution $Q(t;\beta)$ of waiting times between message and action 
for a specific $\beta$, weighted by the population density
$\psi (\beta; \overline{\beta}, \beta_{0})$: 
\begin{equation}
\overline{Q}(t; \overline{\beta}, \beta_{0}) = \int_{-\infty}^{+\infty} Q(t;\beta) \psi(\beta; \overline{\beta}, \beta_{0}) d\beta ~.
\label{cdfwth}
\end{equation}
Using the form of $Q(t;\beta)$ previously obtained \cite{SaiSorProcras}, expression (\ref{cdfwth}) 
predicts different regimes that we now test on our dataset.

\section{Empirical results}

\subsection{Data sets}

We consider a system where word-of-mouth effects and other social interactions are essentially absent, so as 
to test the generalized queueing theory as cleanly as possible. Specifically, 
we study the persistence of the use of outdated Web browsers (Firefox, Opera, Chrome and Safari) after users have been prompted to perform an update. Our data is obtained from anonymized daily log files of Google web servers (more than 70\% of the worldwide daily searches), collected over more than three years \cite{frei2009} (cf. the Appendix). 

The release of a new browser version is typically accompanied by an alert message notifying the user of the pending update. The message is delivered to all users at the same time and the update is performed at different times, if ever, by an heterogenous population of people. 

\subsection{Analysis of data with the priority queuieing model}

Fig. 2 shows the time dependence of the fraction of the population who use a given Mozilla Firefox browser
version. Each time a new version is released, the fraction of users using the outdated version 
drops initially very fast and then very slowly, in favor
of the new version which has its market share increase inversely. Each 
version goes through a life cycle of fast increase of usage followed by a slow
concave asymptotic regime, and then a sharp collapse at the time of the introduction
of the next version, continuing into a slow asymptotic decay.
Almost all versions of all browsers exhibit this characteristic behavior.
  
Fig. 3a shows a few representative decays of browser use after the introduction of a new version, taken
as the origin of time, and their fit by expression (\ref{cdfwth}). For each fit, the corresponding 
distribution $\psi (\beta; \overline{\beta}, \beta_{0})$ is shown in panel 3b (see \cite{allfits} for the fits
of all 44 browser versions that we have analyzed).
\begin{itemize}
\item Mozilla Firefox 3.0.3 (orange circles and line) is well fitted by a simple power law with 
exponent $\alpha \simeq 0.5$, and 
the corresponding probability distribution function  $\psi(\beta; \overline{\beta}, \beta_{0})$ is close to 
a Dirac function centered on $0$. 
\item For Mozilla Firefox 1.5.0.8 (green crosses and line), the decay
is essentially exponential. Its corresponding $\psi(\beta; \overline{\beta}, \beta_{0})$ lies in a narrow range of
negative $\beta$'s ($\overline{\beta} < 0$ and $\beta_0 < |\overline{\beta}|/3$).
This version is associated with a population under utilizing their time and with rapid responses
to perturbations.
\item For Mozilla Firefox 1.5.0.11 (red diamonds and line), the power law decay 
with exponent $\alpha \simeq 0.9 \pm 0.1$ is followed by an upward curvature,
which can be interpreted as the convergence to a non-zero plateau at much larger times.
The corresponding pdf  $\psi(\beta; \overline{\beta}, \beta_{0})$ 
is very broad with $\overline{\beta} <0$ and  $|\overline{\beta}| \simeq \beta_{0}$, 
while the upward curvature is contributed by the fraction of positive $\beta$'s in the population. 
For this browser version, the generally quite fast upgrade is accompanied by 
a large heterogeneity of behaviors, which can be due to complications preventing a 
straightforward execution of the task. 
This version is again associated with a population under utilizing their time but with 
slow and weak adaptation to perturbations.
\item For Apple Safari 2 (SF2),
(blue $+$'s and line), $\overline{Q}(t; \overline{\beta}, \beta_{0})$
decays very slowly as a power law with exponent 
$\alpha \simeq 0.3 \pm 0.1$, which qualifies the procrastination mode discussed with
Fig. 1c. The decay is  followed by an upward curvature, again 
suggesting the convergence to a plateau quantifying the fraction of 
users who will never update their browser. 
\item For SF2, we show two decays following two 
successive announcements for this version update (October 26th and November 16th, 2007).
These two versions are associated with a population over utilizing their time and with 
relatively rapid responses to perturbations.
\end{itemize}

Figure 4 depicts the phase diagram of the different decay regimes in the plane $(\overline{\beta}, \beta_{0})$.
The crosses represent the best parameters fitted to the 44 browser versions updates that
we have analyzed.  The line $\beta_0 = - {1 \over 3} \overline{\beta}$ separates 
a lower region, in which more than $99.5\%$  of $\beta$'s are negative and basically all users
end up executing the target tasks, from an upper domain in which a significant fraction
of the population has $\beta >0$ for which the decay saturates at a plateau giving
the probability for the target task never to be completed.

\section{Conclusion}

We have shown that the generalization of priority queueing models to encompass
a heterogenous and dynamically varying population of users provide a complete rationalization
of the large variability of responses of individuals to messages. These results show that it is possible to measure and predict the efficiency of time allocation by humans, individually or collectively, in a particular or more general context. Our result open new perspective for understanding and modeling decision-making by humans in real life, but also effects in crowds \cite{dhelbing}, in the presence of social interactions and viral epidemics \cite{cranesorYouTube}.

\vskip 1cm
{\bf Acknowledgements}:  We are grateful to Riley Crane for insightful discussions and for providing the basic framework to fit decays. We acknowledge financial support from the ETH Competence Center ÓCoping with Crises in Complex Socio-Economic SystemsÓ (CCSS) through ETH Research
Grant CH1-01-08-2.

\newpage

\section*{Appendix: Data Collection}

We use data obtained by Google Switzerland from anonymized daily log files of Google's global web services including Google search (\url{www.google.com}) performed in two distinct campaigns and over more than three years \cite{frei2009,silent_updates_2009}. We only considered Apple Safari, Mozilla Firefox, Opera and Google Chrome. We had to exclude Microsoft Internet Explorer because we could not retrieve all minor updates and it is used by large organizations (e.g.corporations) that apply specific update policies, which could significantly bias the results.

For each type of browser, we extracted the daily fraction of users for each version of this browser. Since Google searches account for more than 70\% of the hundreds of
millions of searches performed each day on the World Wide Web, our database provides a unique window
to the dynamics of the use of browsers at the largest scale. Data have been collected by Google Inc. in two campaigns over two years:

\begin{itemize}
\item \textbf{Campaign 1} : from 20th December 2006 until 31st July 2008. 
\item \textbf{Campaign 2} : from 21th October 2008 until 17th April 2009. 
\end{itemize}

All browsers daily activity  are recorded either twice a week before 10th October 2007 and after 20th December 2008, and every day between these two dates. 

When a browser communicates with a Web server, it usually sends the user-agent string in the HTTP protocol header with every request for a Web page. This string contains the type and version of the browser and the type of operating system the browser runs on. To ensure that an Internet user visiting several times Google within one day is not counted twice or more, each visit was linked with Google's PREF cookie to eliminate duplicate visits from the same browser. We ignored the small fraction of browsers that disabled cookies due to restrictive user settings. We also ignored the small possibility of cookie id collisions and the effect of users deleting cookies manually. There are also some proxies that change the user-agent string. Based on the observed dynamics of this research, we expect this effect to be small. Some hosts send fake ``HTTP user agent'' string for various reasons (web crawling, bots, etc...), but we can assume that this phenomenon is also marginal compared to the 75\% of the World Wide Web traffic captured by Google at this time \cite{frei2009}.

Google has a strict confidentiality policy to protect users and all data are anonymized after six months. Additionally, we only had access to relative activity of browsers over time per browser. This means that the sum over all versions of each browser (resp. Chrome, Firefox, Opera or Safari) is equal to $1$. Out of 309 browser version traces identified in the database only 44 had fully exploitable decays, namely exhibiting  a fast drop due to an exogenous shock (the release of a newer version) followed by a decay over a sufficient period of time (between 60 and 600 days).

\newpage

\begin{figure}[h]
\centerline{\epsfig{figure=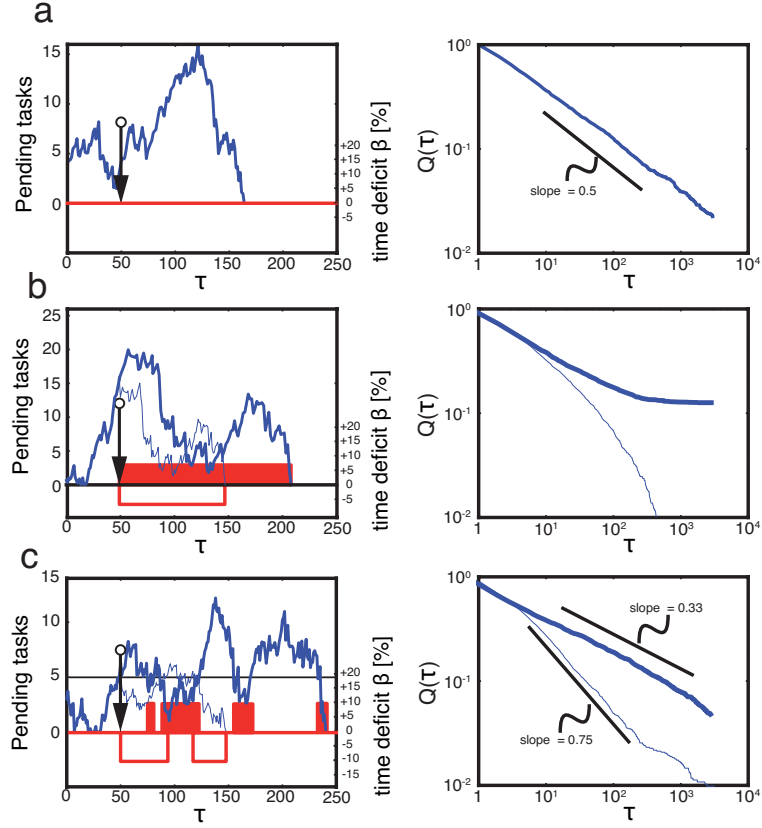,angle=0,width=10cm,scale=1}}
\caption{Simulations of three typical strategies of browsers updates. The left panels show 
typical evolutions of the instantaneous time deficit parameter $\beta$ associated with the queue of pending tasks as a function of time $\tau$, obtained by executing the simple priority queueing model. The arrow indicates the time of arrival of the target task, which is executed when $\beta$ touches $0$. The right panels display the distribution of waiting times $Q(t)$ before the target task 
is executed obtained by simulating a population of $3000$ individuals.
The panels {\bf a}, {\bf b} and {\bf c} correspond to three possible adaptation / reactions of users to 
the incoming target task.
{\bf a}. The arrival of the target task does not perturb the flow
of task execution. {\bf b}. The arrival of the target task triggers a stress (negative or positive); the target task(s) will be executed far much slower (bold line) (resp. faster (light line)) leading to a crossover towards a plateau (resp. exponential decay). {\bf c}. Humans procrastinate, i.e. the target task is not necessarily
executed even in absence of other tasks. The arrival of the target task can in addition 
increase (resp. decrease) the stress and its execution is less likely (bold line) (resp. more likely (light line)).}
\label{fig:model}
\end{figure}

\newpage
\begin{figure}[h]
\centerline{\epsfig{figure=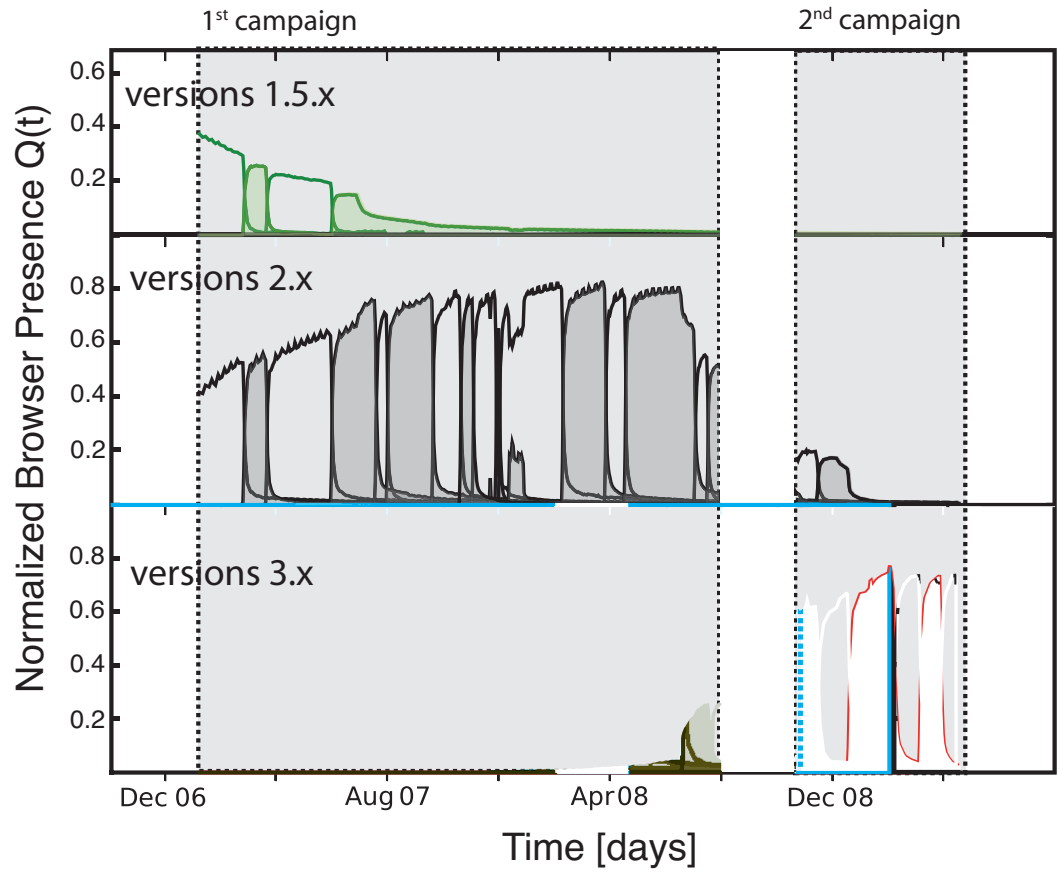,angle=0,width=15cm,scale=1}}
\caption{Fraction of the population using a given Mozilla Firefox browser (minor) version as a function of time and over two data collection campaigns (cf. Appendix). The three major versions 1.5.x, 2.x and 3.x are shown
separately.}
\label{fig:moz_dyns}
\end{figure}

\newpage

\begin{figure}[h]
\centerline{\epsfig{figure=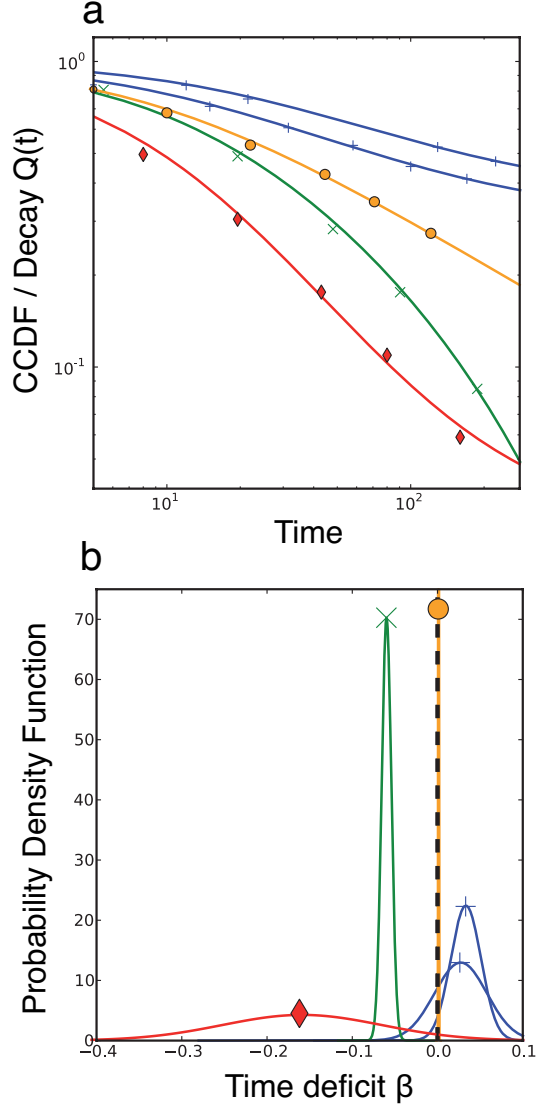,angle=0,width=7cm,scale=1}}
\caption{{\bf a}. Typical decays of the ccdf $\overline{Q}(t; \overline{\beta}, \beta_{0})$
of waiting times till the update to the new version for Mozila Firefox 3.0.3 (green diagonal crosses and line), Mozilla Firefox 1.5.0.8 (yellow circles and line), Mozilla Firefox 1.5.0.11 (red diamonds and line) and Apple Safari 2 characterized by two consecutive decays (blue crosses and lines).(See \cite{allfits} for all 44  fits). Points show the empirical (bined) decays and lines represent the best fit. {\bf b}. Corresponding Gaussian probability distribution function (pdf) $\psi(\beta; \overline{\beta}, \beta_{0})$ of time deficit $\beta$ defined by expression (\protect \ref{gaussian}).}
\label{fig:decays}
\end{figure}

\newpage
\begin{figure}[h]
\centerline{\epsfig{figure=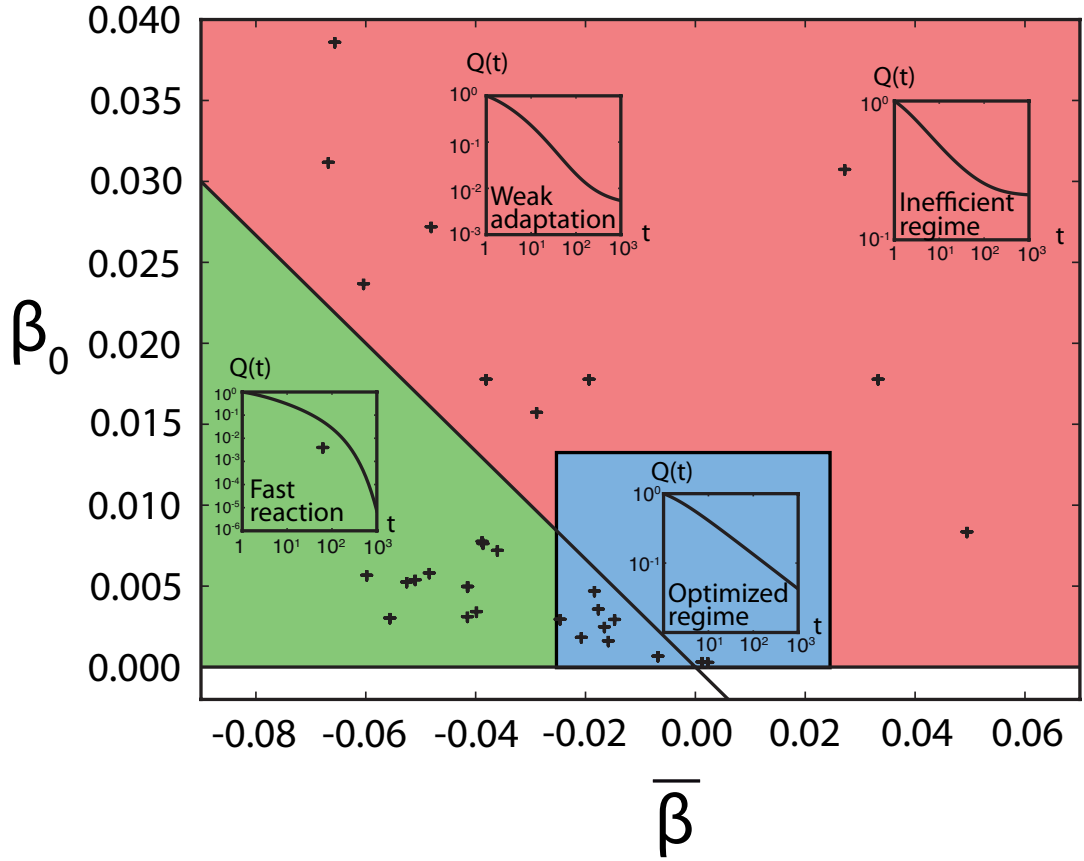,angle=0,width=15cm,scale=1}}
\caption{Phase diagram in the plane  $(\overline{\beta}, \beta_{0})$
summarizing the different regimes for the decay of the ccdf 
$\overline{Q}(t; \overline{\beta}, \beta_{0})$ of waiting times
between message and update, predicted by expression (\protect\ref{cdfwth})
with (\protect\ref{gaussian}). The  44 $+$'s correspond to
the best fits of the 44 data sets with expression (\protect\ref{cdfwth}). 
The line $\beta_0 \approx - {1 \over 3} \overline{\beta}$ separates 
the upper inefficient update region from the lower efficient update region.}
\label{fig:synthes}
\end{figure}

\end{document}